\documentclass{rnaastex}

\begin{document}

\title{An independent confirmation of the future flyby of Gliese 710 to the solar system using {\it Gaia} DR2}

\correspondingauthor{Ra\'ul~de~la~Fuente~Marcos}
\email{rauldelafuentemarcos@ucm.es}

\author[0000-0002-5319-5716]{Ra\'ul~de~la~Fuente~Marcos}
\affiliation{AEGORA Research Group,
             Facultad de Ciencias Matem\'aticas,
             Universidad Complutense de Madrid, 
             Ciudad Universitaria, E-28040 Madrid, Spain}

\author[0000-0003-3894-8609]{Carlos~de~la~Fuente~Marcos}
\affiliation{Universidad Complutense de Madrid,
             Ciudad Universitaria, E-28040 Madrid, Spain}

\keywords{Galaxy: kinematics and dynamics --- stars: individual (WISE~J072003.20-084651.2, Gliese~710)}

\section{} 

Gliese~710 is a K7V star located 19~pc from the Sun in the constellation of Serpens Cauda \citep{2006AJ....132..161G}. It has been known for 
nearly two decades that Gliese~710 is headed straight for the solar system \citep{1999AJ....117.1042G,2002Icar..157..228M,2010AstL...36..220B,
2015MNRAS.454.3267F,2016A&A...595L..10B}. The most recent published analysis, based on {\it Gaia} DR1, concluded that in $1.35\pm0.05$~Myr, 
Gliese~710 will be $13\,366\pm6250$~au from the Sun \citep{2016A&A...595L..10B}. Here, we present an independent confirmation of this 
remarkable result using {\it Gaia} DR2. Our approach is first validated using as test case that of the closest known stellar flyby, by the 
binary WISE~J072003.20-084651.2 or Scholz's star \citep{2015ApJ...800L..17M}.

{\it Gaia} DR2 \citep{2016A&A...595A...1G,2018arXiv180409365G} provides, among other data, right ascension and declination, absolute stellar 
parallax, spectroscopic radial velocity, proper motions in right ascension and declination, and their respective standard errors, all in the 
solar barycentric reference frame. These data can be transformed into equatorial values as described by e.g. \citet{1987AJ.....93..864J}; 
state vectors in the ecliptic and mean equinox of reference epoch suitable for solar system numerical integrations can subsequently be 
computed by applying the usual transformation that involves the obliquity. Using input data from {\it Gaia} DR2 and barycentric Cartesian 
state vectors for the solar system provided by Jet Propulsion Laboratory's \textsc{horizons},\footnote{\url{https://ssd.jpl.nasa.gov/?horizons}} 
we have carried out $N$-body simulations as described by \citet{2012MNRAS.427..728D}. Both input data and integration tools are different 
from those used in previous works.

\begin{figure}[!ht]
\begin{center}
\includegraphics[scale=0.55,angle=0]{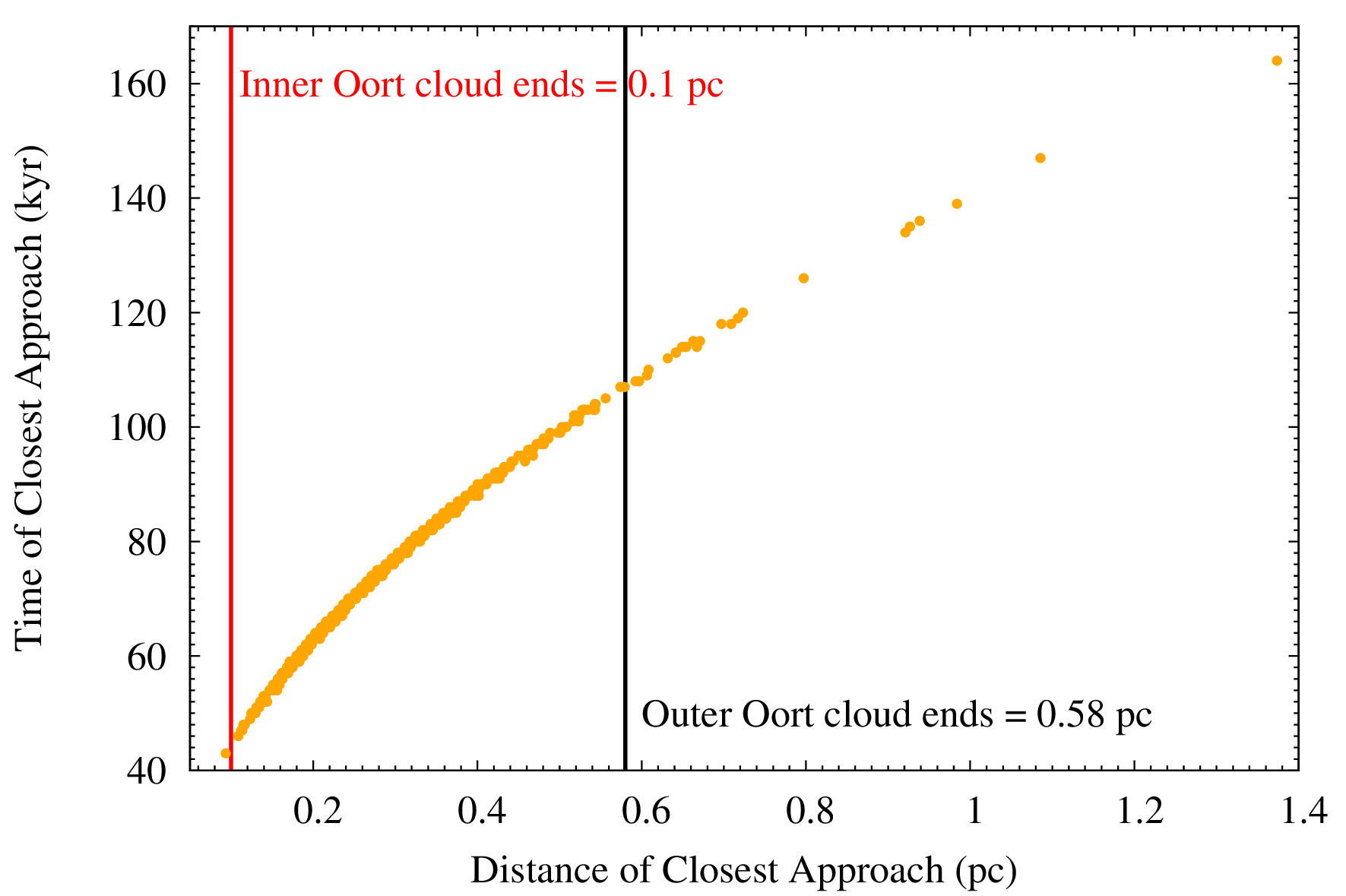}
\includegraphics[scale=0.55,angle=0]{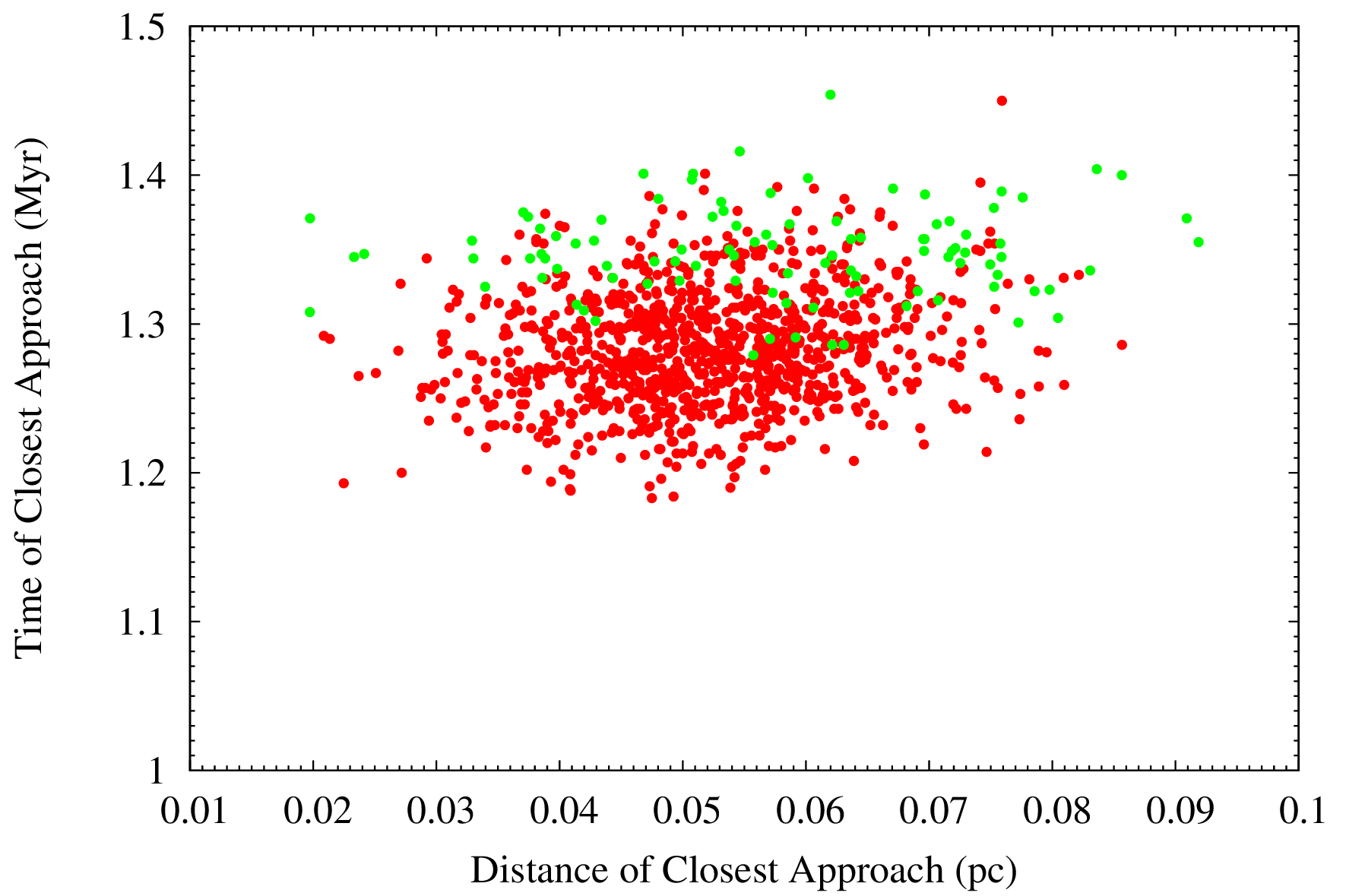}
\caption{Time of closest approach as a function of the distance for the flybys of Scholz's star (top panel) and Gliese~710 (bottom panel) 
         to the solar system (1000 control datasets each). The bottom panel includes results obtained using input data from {\it Gaia} DR2 (red) 
         and older data (green, 100 integrations).
\label{fig:1}}
\end{center}
\end{figure}

For validation purposes, we have used the flyby of Scholz's star investigated by \citet{2015ApJ...800L..17M} as test case; this work states 
that WISE~J072003.20-084651.2 may have passed $0.25^{+0.11}_{-0.07}$~pc from the Sun, $70^{+15}_{-10}$~kyr ago. We have used the same input 
data (Scholz's star has no public data in {\it Gaia} DR2) and our independent approach to compute the evolution backward in time of 1000 
control datasets of this star. Figure~\ref{fig:1}, top panel, shows our results: the perihelion distance is $r=0.28\pm0.12$~pc and it 
happened $t_{r}=73\pm14$~kyr ago (averages and standard deviations, the median and interquartile range, IQR, values are $0.25\pm0.12$~pc and 
$71\pm16$~kyr, respectively), the closest approach might have reached 0.09~pc or 19\,312~au, 43~kyr ago. Our results are consistent with 
those in \citet{2015ApJ...800L..17M}. Applying identical methodology to Gliese~710, but this time using input data from {\it Gaia} DR2 
(object {\it Gaia} DR2 4270814637616488064) and integrations forward in time, we obtain Figure~\ref{fig:1}, bottom panel: now $r=0.052\pm0.010$~pc 
(or $10\,721\pm2114$~au) and $t_{r}=1.28\pm0.04$~Myr into the future (averages and standard deviations, the median and IQR values are 
$0.052\pm0.013$~pc and $1.28\pm0.05$~Myr, respectively), the closest approach might reach 0.021~pc or 4303~au, in 1.29~Myr (using older input 
data, we obtain $r=0.06\pm0.02$~pc and $t_{r}=1.35\pm0.03$~Myr). 

Our results confirm, within errors, those in \citet{2016A&A...595L..10B}, but suggest a closer, both in terms of distance and time, flyby of
Gliese~710 to the solar system. Such an interaction might not significantly affect the region inside 40~au as the gravitational coupling 
among the known planets against external perturbation can absorb efficiently such a perturbation \citep{1997AJ....113.1915I,2007PASJ...59..989T}, 
but it may trigger a major comet shower that will affect the inner solar system (see e.g. \citealt{2015ApJ...800L..17M,2016A&A...595L..10B,
2017MNRAS.472.4634K,2018MNRAS.476L...1D}). 

\acknowledgments

We thank S.~J. Aarseth for providing the code used in this research and A.~I. G\'omez de Castro for providing access to computing facilities. 
This work was partially supported by the Spanish MINECO under grant ESP2015-68908-R. In preparation of this Note, we made use of the NASA 
Astrophysics Data System. This work has made use of data from the European Space Agency (ESA) mission {\it Gaia} 
(\url{https://www.cosmos.esa.int/gaia}), processed by the {\it Gaia} Data Processing and Analysis Consortium (DPAC, 
\url{https://www.cosmos.esa.int/web/gaia/dpac/consortium}). Funding for the DPAC has been provided by national institutions, in particular 
the institutions participating in the {\it Gaia} Multilateral Agreement.

\end{document}